# ARTICLE

# Effects of withdrawal speeds on the structural, morphological, electrical, and optical properties of CuO thin films synthesized by dip-coating for $CO_2$ gas sensing

A.M.M. Musa[a,b], S.F.U. Farhad[*a,c], M.A. Gafur[a] and A.T.M.K. Jamil[b]

Copper oxide (CuO) thin films have been deposited on glass substrates by a facile sol-gel dip-coating technique with varying withdrawal speeds from 0.73 to 4.17 mm/s. The variation of film thickness manifested by dip-coating withdrawal speeds was investigated in detail to investigate its effect on the structural, morphological, opto-electrical, and wettability properties of CuO thin films for carbon dioxide ($CO_2$) gas-sensing applications. The crystallinity, as well as phase purity of dip-coated CuO, were confirmed by both X-ray diffraction (XRD) and Raman spectral analyses. The surface morphology of the films characterized by the scanning electron microscopy (SEM) revealed that pore density is decreasing with increasing withdrawal speeds and grain size is found to increase with increasing film thickness corroborating the XRD results. The optical bandgap of dip-coated CuO films was estimated in the range of 1.47-1.52 eV from the UV-VIS-NIR transmission data and it is found decreasing with the increment of Urbach tail states accompanied by the increase of film thickness. The ratio of the electrical and optical conductivity of CuO films was found to decrease with increasing withdrawal speeds due to the variation carrier concentration. Among all the studied films, the sample deposited at 0.73 mm/s withdrawal speed exhibited the highest crystallinity, porous morphology, pore density, opto-electrical conductivity as well as large water contact angle, and therefore maximum gas sensing response of $CO_2$ vapor in the air recorded at room temperature.

## 1. Introduction

Copper forms disparate oxides due to its polyvalency, among them, well-known cupric oxide (CuO) and cuprous oxide ($Cu_2O$) are generally p-type semiconductors due to the copper vacancies ($V_{cu}$) in the crystal lattice.[1,2] Cupric oxide (CuO) also known as tenorite is the most stable phase of the copper oxides family and it shows high optical absorption, reported narrow band gap energy between 1.2 to 2.1 eV, along with the ability to synthesis in different nanostructured morphologies.[3,4] Due to its abundance in earth crust, environmentally benign nature as well as tailorable electro-optical properties, recently CuO attracts tremendous attention in the thin-film technology to explore its potential device application such as solar cells,[5] electrochemical devices,[6] and more importantly for gas sensors.[1,7,8,9] In the case of gas sensing, CuO thin film-based sensors are particularly alluring due to their chemical stability, large water contact angle (>90°), high sensitivity, faster response time, lower operating temperature as well as proven ability to detect diverse gases such as $NO_2$, $H_2$, CO, $CO_2$, etc.[7,8,10] Carbon dioxide ($CO_2$) has been extensively used as a refrigerant in fire extinguishers, life jackets also escalated life rafts, plastic, food packaging as well as in carbonated drinks, etc.[9] Therefore, reliable, selective, and fast detection of carbon dioxide gas appears to be the strict necessity for various industrial purposes.[5] However, most of the $CO_2$ sensing devices are bulky, suffer from short-life time, expensive maintenance as well as their vulnerability to interference gases.[7]

Therefore, for $CO_2$ sensing applications, CuO thin films are the most widely studied because of its attractive features mentioned above. However, the gas sensing mechanism is also related to various structural parameters (crystallite size and dislocation densities), the thickness of the film as well as its morphological properties. The change in resistance of materials originates from the surface reaction with ambient gaseous species in addition to the material compositions. To this end, the porous structure of CuO film may be beneficial for the surface reaction along with gas diffusion leading to a faster sensing response.[11] These particularities of CuO make it a strong candidate in $CO_2$ sensing field.

[a] Bangladesh Council of Scientific and Industrial Research (BCSIR), Dhaka 1205, Bangladesh
[b] Department of Physics, Dhaka University of Engineering & Technology (DUET), Gazipur-1707, Bangladesh
[c] Central Analytical and Research Facilities (CARF), BCSIR, Dhaka 1205, Bangladesh
*Correspondence: sf1878@my.bristol.ac.uk ; s.f.u.farhad@bcsir.gov.bd
Electronic supplementary information (ESI) available. See DOI: xxxxxx

CuO thin films can be grown by different deposition techniques such as thermal oxidation,[12] thermal evaporation,[13] filtered cathodic vacuum arc,[14] pulsed laser deposition,[15] Sputtering,[6] electro-spinning,[16] microwave hydrothermal synthesis,[17] the hydrothermal method,[18] spray pyrolysis,[19] and diverse sol-gel based synthesis techniques[7,8,20,21] etc. Among them, the sol-gel dip-coating technique is the simplest and most economical one to deposit thin film under a non-vacuum environment with good control over growth parameters preferably at low processing temperatures.[22,23] In this addition, the number of dip-cycles as well as withdrawal speeds can be influential factors for developing thin films with desired features. In this regard, the physical characteristics of samples can be customized by the variation of film thickness manifested by dip-coating withdrawal speeds and dip-cycles[24]. There are some reports on the influence of various preparation techniques and deposition of CuO thin films,[25] but the effects of the film thickness to tailor the physical properties for $CO_2$ gas sensing applications of CuO deposited by the dip-coating technique are rarely explored. It is also appeared in the literature that most of the $CO_2$ gas sensing works were conducted at elevated temperatures (>200 $^0$C). For examples, Umar et al.[18] reported that CuO nanoplates-based sensor for ethanol, carbon dioxide, other volatile gas detection operated at 250 $^0$C. Bhowmick et. al.[7] reported spin-coated CuO thin film with varying thickness for detecting volatile organic compounds (VOCs) operated at 200 - 300 °C and obtained maximum response for alcohol and acetone at 300 °C for 240 nm thick CuO films. Chapelle et al.[26] reported RF- sputtered CuO/$CuFe_2O_4$ thin films and achieved the best $CO_2$ sensing response at an optimal operating temperature of 250 °C. They have demonstrated the highest response to 5000 ppm of $CO_2$ by adopting a highly porous CuO thin layer atop the $CuFe_2O_4$ film. Recently, Abdelmounaïm et. al.[19] reported copper oxide thin films deposited by spray pyrolysis by varying molar centration of precursor solutions and they have only demonstrated $CO_2$ gas response of phase pure CuO porous thin film. The thin film-based gas sensors' performance chiefly depends on the microstructural, morphological, and physicochemical properties (e.g., hydrophobicity etc.) of the film materials, and these properties can be tailored by optimizing the processing parameters of any suitable synthesis techniques. In the present study, we investigated the effect of withdrawal speeds on the physical properties of CuO thin films prepared by the sol-gel dip-coating technique and finally structural, optical, and morphological properties were analysed in detail for investigating the suitability of dip-coated CuO films for $CO_2$ gas sensing applications at room temperature. Here, we made an attempt to correlate experimental results with the light of possible mechanisms underlying the phenomena.

## 2. Experimental

### 2.1. Substrate preparation

Prior to thin film deposition, ordinary soda-lime glass (SLG) microscopy slides were initially scoured with detergent soap in clean tap water then immersed into a hot water-filled beaker for ultra-sonic cleaning. To remove the grease and other stubborn dirt from the SLG surface, they were further subjected to sequential ultrasonic cleaning in methanol, acetone, ethanol, and lastly in deionized (DI) water at 50 °C ambient, each step for 20 min. Finally, to eliminate moisture from the surface of the substrate, they were first dried by a dry nitrogen gun then by air annealing at 100 °C for 5 min.

### 2.2. Thin film fabrication

Copper acetate monohydrate [$Cu(CH_3COO)_2H_2O$] (purity~99.99%, MERCK Germany) was used as a starting material for preparing the precursor solution. Diethanolamine (DEA) [$C_4H_{11}NO_2$] (purity~99.99%), isopropanol (IPA) [$C_3H_7OH$] (purity~99.99%) and ethylene glycol [$C_6H_6O_2$] (purity~99.99%) Gysmecol (Guangzhou) Technology Co. Ltd. China) were used respectively as the stabilizer, solvent, and coating materials (Gysmecol (Guangzhou) Technology Co. Ltd. China). All chemicals were collected from the local market and used without further purification.

Briefly, copper acetate powder (molar mass: 199.65 g/mol) of 0.699 g was dissolved with 9 ml isopropanol and 0.5 ml DEA for preparing 0.35 M precursor solution. Then 0.5 ml ethylene glycol was added to this precursor solution to obtain 10 ml of solution. After that, this solution was stirred at 500 rpm with a magnetic bar for 120 min at room temperature using a magnetic stirrer. The colour of the final solution was dark blue and clear without any suspension of particulates. The solution was then filtered using a 0.45 µm polytetrafluoroethylene (PTFE) filter before dipping the glass substrates into it. Dip-coating was carried out using a commercial dip coater (model: PTL MM01, OPTOSENSE, UK) with a withdrawal speed of 0.73, 1.75, 3.39, and 4.17 mm/s. After deposition, CuO thin films were immediately dried at 300 °C for 30 min in a furnace (model: CWF 11/13, CARBOLITE GERO, UK) with a heating rate of 5˚ C/min. This process was repeated five times to get five layers for each sample to achieve the desired film thicknesses. Finally, the deposited films were annealed 5 min at 600 °C with the same heating rate for re-crystallization.

### 2.3. Thin film characterizations

The structural characterization of the deposited thin films was performed by X-ray diffraction (XRD) technique using a Bruker D8 Advance Diffractometer (Germany) with a monochromatic copper anode source of Cu-K$_\alpha$ radiation, λ = 1.5406 Å in out-of-plane geometry. The vibrational structure was recorded at room temperature in the range of 100 -700 cm$^{-1}$ by a Raman spectrometer (Horiba Macro Raman) using laser excitation wavelength 785 nm and laser power ≤ 5 mW (laser spot size ~ 0.5 mm). The optical transmission data were recorded by a UV-VIS-NIR spectrophotometer (SHIMADZU -2600, Japan) in the wavelength range of 200-1100 nm at room temperature. The film thickness was measured by an Alpha-Step D-500 stylus profilometer. The surface morphology of deposited films was investigated by using a scanning electron microscope (SEM, ZEISS EVO18). The water contact angle (WCA) of the dip-coated thin films were measured using a contact angle goniometer coupled with a high-resolution video camera (Ossila UK Ltd.) and measurements were taken at three different locations of each film. The elemental composition of dip-coated CuO thin films were investigated by a wavelength dispersive X-ray Fluorescence (WDXRF) spectrometer (Rigaku ZSX Primus IV). The p-type of conductivity of dip-coated CuO films was confirmed by a simple hot point probe technique. The $CO_2$ gas sensing properties of thin films were investigated at room temperature by a computerized gas sensor system coupled with an Electrometer (Keithley 6517B) and the concentrations of $CO_2$ gas were controlled using a digital gas flow meter (Model: PH-1004, Henan, China).

## 3. Results and discussion

### 3.1. Structural analyses

XRD patterns of the thin films synthesized with different dip-coating withdrawal speeds are compared in Fig. 1a. CuO films exhibit well defined diffraction peaks at 2θ = 32.52, 35.58, 38.7, 48.84, 53.48, 58.24, 61.7, 66.28, 68.08 and 72.4°, respectively assignable to (110), ($\bar{1}$11), (111), ($\bar{2}$02), (020), (202), ($\bar{1}$13), ($\bar{3}$11), (220) and (311) planes of monoclinic CuO phase (ICDD card no. 48-1548). These diffractograms also suggest a polycrystalline nature of the CuO films with two prominent peaks at 32.52° and 35.58° corresponding to ($\bar{1}$11) and (111) planes.[19] No Bragg's peaks corresponding to $Cu_2O$ and/or metallic Cu phase are discernible in the XRD patterns suggesting that our dip-coated thin films are basically composed of phase pure CuO. Room temperature Raman spectra were also recorded to further confirm the phase purity of these films and shown in Fig. 1b.

Figure 1b compares the dip-coated thin films with varying withdrawal speeds. The major phonon modes reported are ~296 cm$^{-1}$($A_g$), ~346 cm$^{-1}$ ($B_g^{(1)}$), and ~630 cm$^{-1}$ ($B_g^{(2)}$) for CuO phase and ~150 cm$^{-1}$($T_{1u}^{(1)}$), ~220 cm$^{-1}$ ($2E_u$), ~304 cm$^{-1}$ ($A_{2u}$), ~520 cm$^{-1}$ ($T_{2g}$) ~615 cm$^{-1}$ ($T_{1u}^{(2)}$) for $Cu_2O$ phase.[2,14,15] Clearly, Raman peaks of the dip-coated thin films are in good agreement with the monoclinic CuO phase and free of other copper-oxide phases which corroborate the XRD results. The intensity of major Raman peak $A_g$ of CuO for 0.73 mm/s dip-coated films is comparatively pronounced than other films suggesting its improved crystallinity as well as highest thickness among all dip-coated films studied herein. This is also consistent with the measured film thickness shown in Fig. 2.

The estimated film thickness as a function of the dip-coating withdrawal speeds is shown in Fig. 2. A growth rate of (-23.92 ± 3.49) nm/mms$^{-1}$ can be determined from a linear fit of the data points. This study suggests that deposited film thickness is inversely proportional to the dip-coated withdrawal speed.[27,28] Therefore, this growth rate may be utilized for achieving desired thickness of CuO dip-coated films for various applications.

The microstructural parameters such as crystallite size *(D)*, micro-strain *(ε)*, and dislocation density *(δ)* were calculated using the following equations:[29]

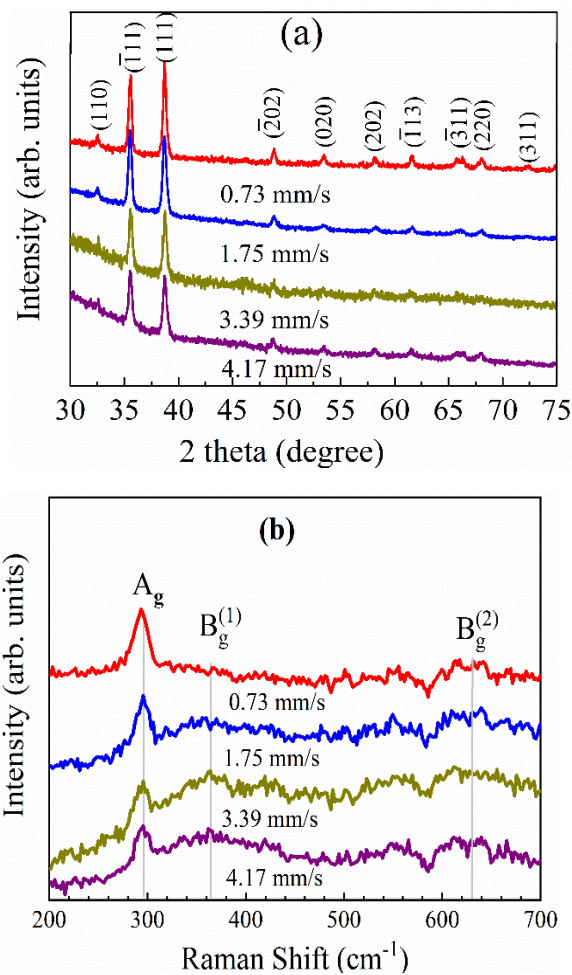

Fig. 1 (a) XRD patterns and (b) Raman spectra of CuO thin films deposited by different withdrawal speeds. The vertical lines in FIG. 1b represent the vibrational peaks of phase pure CuO.[2,15]

$$D = \frac{k\lambda}{\beta cos\theta} \quad (1)$$

$$\varepsilon = \frac{\beta cos\theta}{4} \quad (2)$$

$$\delta = \frac{1}{D^2} \quad (3)$$

Where *k* is a constant usually depends on the shape of the crystallite (*k* = 0.94 for spherical crystallites or grains), *λ* is X-ray wavelength, *β* is the full with at half maximum (FWHM) in degrees which are corrected in case of instrumental broadening estimated by fitting the peaks using XRD tool, and *ϑ* is the Bragg angle of the *{h k l}* reflection.

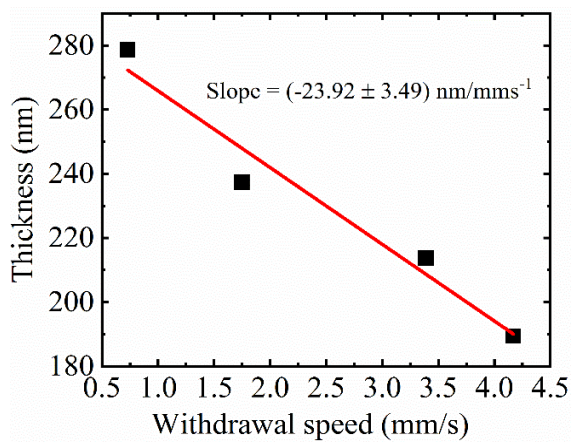

Fig. 2. Thickness variation of CuO thin film as a function of dip-coating withdrawal speeds. The red line in the graph shows the linear fit of the data points.

optimal deposition condition. The increment of crystallite size is consistently associated with strain value and indicates a reduction in the concentration of lattice imperfections. Meanwhile, the lattice parameter $b$ remains constant up to 238 nm film thickness then it decreases with increasing film thickness.[34]

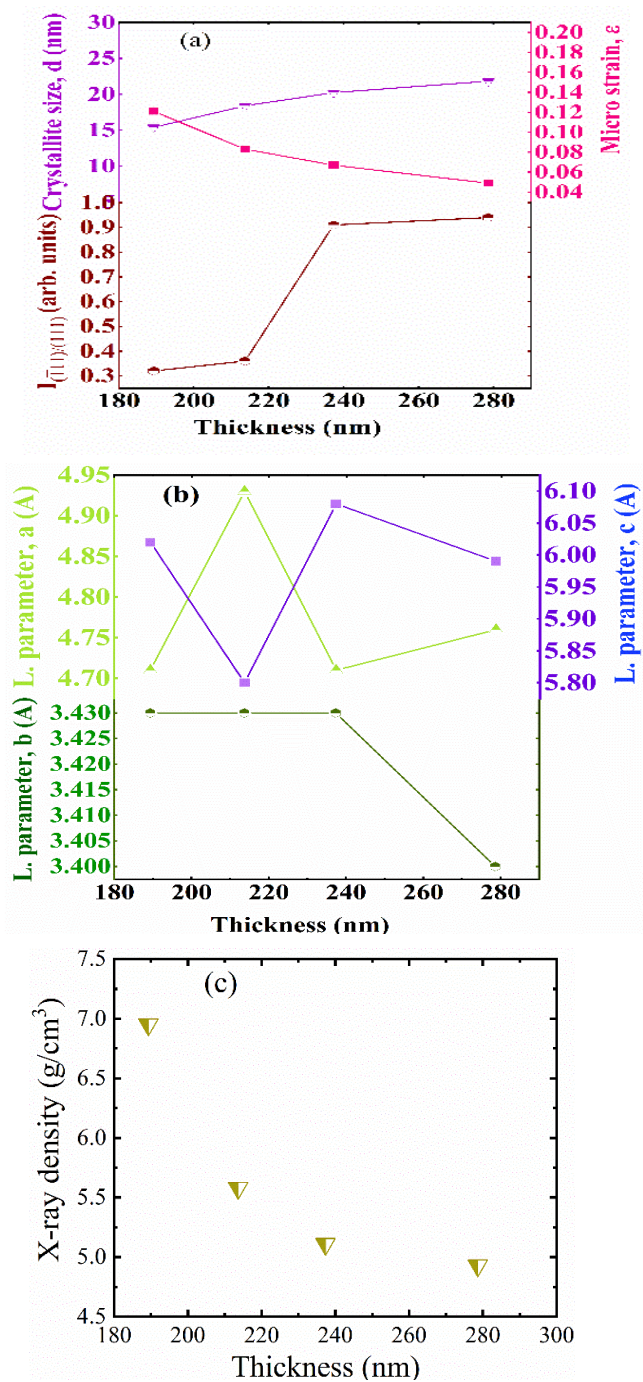

The variation of important structural parameters as a function of the average film thickness was estimated using equation (1)-(3) and depicted in Fig. 3.

Figure 3a shows the variation of crystallite size *(D)* estimated from predominant (111) reflection peak, the intensity ratio of $(\bar{1}11)$ and (111) reflections, and the micro-strain, $\varepsilon$ as a function of film thickness and important parameters are also summarized in Table 1. It is observed that crystallite size increases from 15.41 to 21.83 nm with an increase of film thickness from 189 to 279 nm as expected. The thicker films possess larger crystallite sizes associated with smaller dislocation densities and smaller strains consistently; suggesting the improvement of the films' crystallinity, a decrease of lattice imperfection with an increase of film thickness.[19] In other words, the decrease of the crystal strain reflects the stress relaxation[30] in the dip-coated thicker CuO films.

The formula used to calculate the lattice parameters (*a*, *b*, and *c*) from the predominant (-111) diffraction peak of monoclinic CuO thin films are given by:[31]

$$\frac{1}{d^2} = \frac{1}{\sin^2\beta}\left(\frac{h^2}{a^2} + \frac{k^2 \sin^2\beta}{b^2} + \frac{l^2}{c^2} - \frac{2hl\cos\beta}{ac}\right) \quad (4)$$

The angle $\gamma$ is 99.54° for monoclinic structure. The standard values of lattice parameters are, *a* = 4.6837 Å, *b* = 3.4226 Å, *c* = 5.1288 Å and ratios are, *a/b* = 1.3684 and *c/b* = 1.4985 indexed with ICDD card no. 48-1548. The standard and calculated values of lattice constants (see Fig. 3b) and are in good compliance with those reported in the literature.[31,32] Figure 3b shows that the lattice parameters *a* is inversely related with *c*. As a function of film thickness, when parameter *a* increases the parameter *c* decreases, and vice-versa. This may occur due to the strain value is significantly lower because of the residual stress in the deposited films.[33] Therefore, residual stress in the thickest film deposited at 0.73 mm/s withdrawal speed should be the lowest accompanied by the lowest micro strain; this may also suggest a more stoichiometric CuO film achieved at this

Fig. 3 (a) Variation of the average intensity $I(\bar{1}11)/I(111)$ ratios, crystal size, and micro strain, (b) variation of lattice parameters *a*, *b* and *c*, and (c) variation of X-ray volume density as a function of the film thickness of dip-coated CuO films.

Table 1 Structural parameters associated with $(\bar{1}11)$ diffraction plane with different dip-coating withdrawal speeds

| Withdrawal speed (mm/s) | Thickness (nm) ± 5 | 2θ (deg.) | FWHM β (deg.) | Crystallite size, D (nm) ± 0.01 | Dislocation Density, δ (lines/nm$^2$) | Micro strain, ε × (10$^{-3}$) | X-Ray density, ρ (g/cm$^3$) |
|---|---|---|---|---|---|---|---|
| 0.73 | 279 | 35.56 | 0.40 | 21.83 | 2.09 | 0.096 | 4.93 |
| 1.75 | 237 | 35.59 | 0.43 | 20.23 | 2.44 | 0.010 | 5.11 |
| 3.39 | 214 | 35.60 | 0.48 | 18.40 | 2.95 | 0.104 | 5.58 |
| 4.17 | 189 | 35.61 | 0.54 | 15.41 | 4.21 | 0.121 | 6.95 |

Figure 3c shows the X-ray volume density of these films is calculated using the following formula:[33]

$$\rho = \frac{nM_w}{NV} \qquad (5)$$

Where $n$ is the number of atoms per unit cell, $M_w$ is the molecular weight of CuO, V is the volume of the unit cell and $N$ is Avogadro's number.

Notice that the improvement of films' crystallinity manifested by the increase of number of crystallites per unit volume of CuO thin film. This result confirms that when the thickness was increased the crystal growth process was improved, which intern leads to a decrease in the crystal defects of CuO films (see the 2$^{nd}$ last column of Table 1). The estimated values of X-ray volume densities are listed in the far-right column of Table 1. It is seen that when CuO film thickness increases from 189 nm to 279 nm, the estimated values of X-ray volume density also decrease from 6.95 to 4.93 g/cm$^3$. The X-ray density depends on how much of the X-ray beam absorbed in the sample. The dense film absorbs more X-ray beam compared to the lesser dense structures. In another words, the less dense structure may accumulate more porosity in deposited film (see also Fig. 4 below). In lower withdrawal speed, there may be more and more adsorbed species on the substrate surface and associated solvents impregnated into the layer compared to those grown using higher withdrawal speeds. Therefore, when annealing them, it produces more porosity in the thicker films compared to thinner films as higher withdrawal speed yields the thinner films with comparatively less solvent impregnated into it. This effect should be seen in the SEM surface morphological and optical analyses as discussed in section 3.2 and 3.3 below.

### 3.2. Morphological analyses

The SEM micrographs of CuO thin films dip-coated by using 0.73, 1.75, 3.39, and 4.17 mm/s withdrawal speeds are shown in Fig. 4 (a-d). It is conspicuous that films grown with 0.73 mm/s withdrawal speeds has the highest porous microstructure compared to the other films and their porous nature seems to disappear for films grown with increasing withdrawal speeds (cf. Fig. 4b - Fig. 4d). The high number of pores in the film surface should increase the accessibility of the sensing gas to the bulk of the film thereby enhance the available active area for gas adsorption. The particles size was estimated using ImageJ software and it was found to be 429 ± 65 nm for the sample deposited at 0.73 mm/s, 366 ± 170 nm at 1.75 mm/s, 248 ± 21 nm at 3.39 mm/s, and 241 ± 50 nm at 4.17 mm/s. This observation of grain size variation is consistent with the crystallite size estimated from XRD analyses. It can be seen that thicker film possess larger size grains which may increase the porous nature of the deposited film during thermal treatment process by escaping organic solvent inside the film. The bulk porosity of dip-coated films was also investigated by means of optical studies in details and discussed in the section 3.3 below.

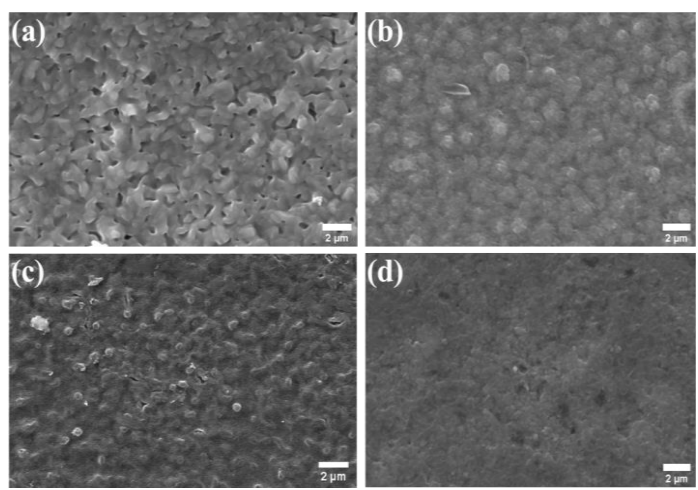

Fig. 4 SEM micrographs of representative CuO thin films dip-coated by using (a) 0.73, (b) 1.75, (c) 3.39, and (d) 4.17 mm/s withdrawal speeds.

### 3.3. Optical studies

The optical transmission *(T)* of the films with varying withdrawal speeds and the plot of ln *(αhυ)* versus *ln (hυ - $E_g$)* for investigating the transition nature of the optical bandgap ($E_g$) of the CuO film as shown in Fig. 5a and 5b respectively. It is seen that the mean transmission of the films decreases from 70 to 33% in the range 800 to 1100 nm (see the inset of Fig. 5a) with increasing film thickness.

The value of *T* increases sharply with the increase of wavelength in the visible region and is found to be transparent in the NIR region of spectra. Moreover, decreasing optical transmittance with withdrawal speeds can be attributed to the increasing film thicknesses (see also Fig.3.).[35] The absorption coefficient *(α)* of all samples can be determined using the following relation:[29]

$$\alpha = \frac{1}{t} \ln\left(\frac{1}{T}\right) \quad (6)$$

Here, *t* and *T* represent the film thickness and the percentage of transmittance respectively. The relationship between the optical band gap ($E_g$) and absorption coefficient (*α*) may be expressed as:[36]

$$\alpha h\upsilon = A(h\upsilon - E_g)^m \quad (7)$$

Where *A* is a constant depends on the nature of the transition occurring in the film indicated by various values of *m* related to transitions involved, *h* is a Plank's constant, and *υ* is the illuminated radiation photon frequency. Therefore, equation (7) can be written as,

$$\frac{d(\ln\alpha h\upsilon)}{d(h\upsilon)} = \frac{m}{h\upsilon - E_g} \quad (8)$$

Plot of $d(\ln\alpha h\upsilon)/d(h\upsilon)$ versus $h\upsilon$ indicates diverge at $h\upsilon = E_g$ which get a rough estimation about $E_g$. The experimental value of *m* could be obtained from the slope of the $\ln(\alpha h\upsilon)$ versus *ln (hυ - $E_g$)* plot as shown in Fig. 5b.

The calculated values of *m* of deposited CuO thin films are: 0.49, 0.47, 0.44 and 0.45 related with withdrawal speeds 0.73, 1.75, 3.39, and 4.17 mm/s respectively, which are close to the *m* = 0.5, indicating the direct allowed transition of these films.

In Fig. 6a the optical transmission data were utilized to generate the so-called Tauc plot. The estimated values of $E_g$ are found by the extrapolation of the linear part *(αhυ)²* to the X-axis and summarized in Table 2. The optical bandgap of the dip-coated films is found to be 1.47, 1.48, 1.49, and 1.52 eV for withdrawal speeds 0.73, 1.75, 3.39, and 4.17 mm/s respectively (see also Fig. S1 in the supplemental file) and they are closely matched with the optical bandgap of CuO thin films reported in the literature.[5,24] The band gap energy decreases with increasing film thickness due to the increment of crystallite size [37] as well as possible tail states near the band edges of CuO (cf. Table 1 and Fig. 4).

The Urbach energy tail can be estimated by the following equation:[38]

$$\alpha = \alpha_0 \exp\left(\frac{h\upsilon}{E_u}\right) \quad (9)$$

Where, $\alpha_0$ is a constant and $E_u$ denote the Urbach energy called Urbach tail. Equation (9) can be written as *lnα = (1/$E_u$) (hυ) + ln$\alpha_0$*. The variation of absorption coefficient (*α*) with wavelength of photons for all films can be found in Fig. S1b. The Urbach tail can be expressed as the reciprocal of the slope of the linear part of *lnα* versus *hυ* plot which is shown in Fig. 6b (see also Fig. S4 for the whole range of *lnα* versus *hυ* plot in the supplemental file). It is observed that with the increase in dip-coating withdrawal speeds the Urbach energy tail decreases and these values are listed in Table 1. It is clearly evident that narrowing of the optical bang gap results from the extension in the band tail (see 2nd last column in Table 2) may be due to either higher disorder of phonon states or higher carrier (doping) concentrations[39] in the dip-coated CuO film associated with the increase of film thickness. It is reasonable to infer that higher doping concentration would increase the optical as well electrical conductivity of the associated films.

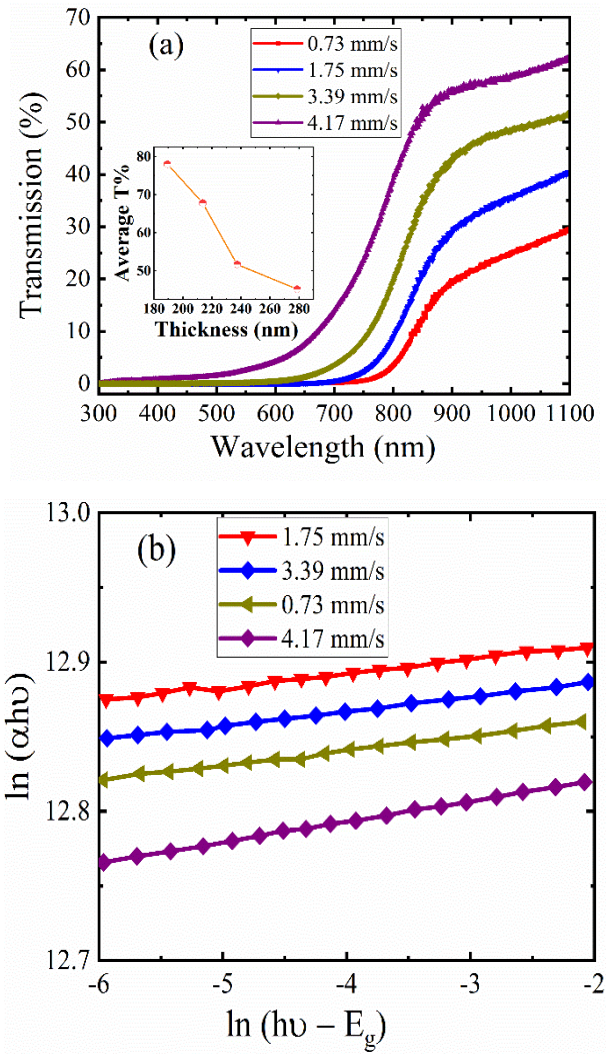

Fig. 5 (a) UV-visible transmission spectra of the deposited CuO thin films. (Inset: Variation of mean transmittance as a function of film thickness), (b) Plot of *ln(αhυ)* versus *ln(hυ - $E_g$)*.

# ARTICLE

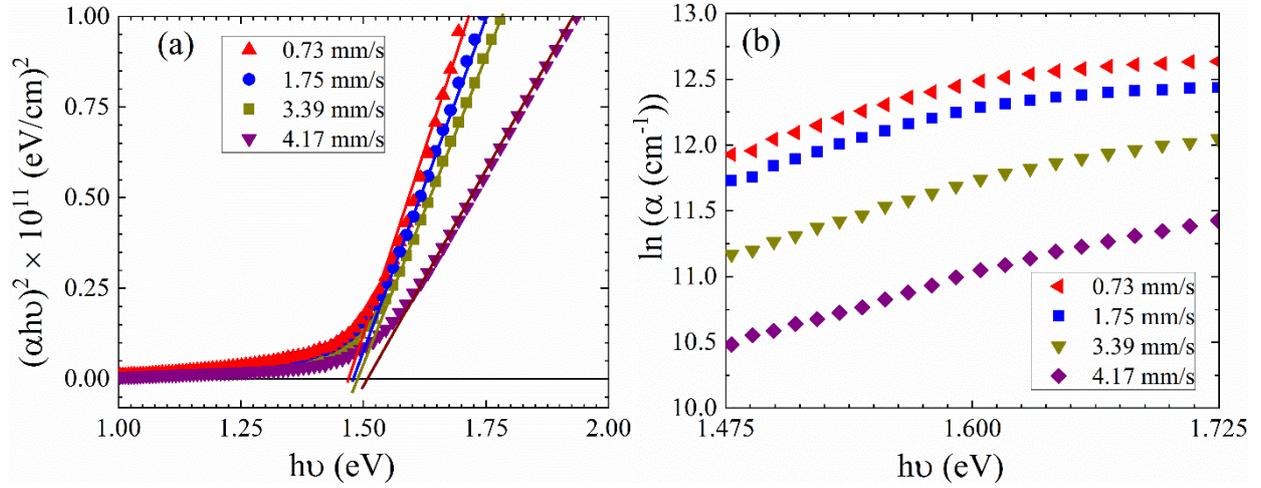

Fig. 6 (a) *(αhυ)²* versus *hυ* plot and (b) *lnα* versus *hυ* plot of CuO thin films synthesized by using various withdrawal speeds for extracting the optical band gap and Urbach tail energy respectively.

Table 2 Optical bandgap and Urbach tail of dip-coated CuO films with the variation of withdrawal speeds

| Molar concentration (M) | Withdrawal speed (mm/s) | Thickness (nm) ± 5 | Grain size_SEM (nm) | Crystallite size_XRD, $D$ (nm) ± 0.01 | Direct band gap, $E_g$ (eV) | Urbach tail, $E_u$ (eV) | Optical conductivity ($S^{-1}$) ($hυ$ = 2.0 eV) |
|---|---|---|---|---|---|---|---|
| | 0.73 | 279 | 429 ± 65 | 21.83 | 1.47 | 0.245 | 7.8×10¹⁵ |
| 0.35 | 1.75 | 237 | 366 ± 170 | 20.23 | 1.48 | 0.234 | 6.9×10¹⁵ |
| | 3.39 | 214 | 248 ± 21 | 18.40 | 1.49 | 0.225 | 5.1×10¹⁵ |
| | 4.17 | 189 | 241 ± 50 | 15.41 | 1.52 | 0.221 | 4.5×10¹⁵ |

### 3.3.1. Optical conductivity ($\sigma_{op}$)

The response of the material to electromagnetic waves is an optical phenomenon. Optical conductivity ($\sigma_{op}$) measurement is one of the most powerful methods on account of proving the progress of quasi-particle excitation and the size of the energy gap in a (super)conductor. Optical conductivity is determined by using the following relation:[40]

$$\sigma_{op} = \frac{\alpha n c}{4\pi} \quad (10)$$

Where $\alpha$ is the absorption coefficient, $n$ is the refractive index, and $c$ is the velocity of light. Figure 7a shows the optical conductivity of CuO thin films associated with thickness. It can be observed that the value of $\sigma_{op}$ rises due to the increment of film thickness and the film prepared to 0.73 mm/s exhibited the highest conductivity of 7.8×10¹⁵ $S^{-1}$. The increase in $\sigma_{op}$ with the increase of film thickness is due to the high absorbance of CuO thin films manifested by the reduction of crystallite defects in CuO films (see Table 1). Higher value of $\sigma_{op}$ for CuO thin film with increasing thickness feature is desirable for its potential integration in optoelectronic devices.

Figure 7b illustrates the variation of band gap energy of CuO films as a function of thickness. This graph implies that the optical band gap of dip-coated CuO decreases with increasing the film thickness. On the other hand, we observed that the optical conductivity rises with increasing thickness of the films. With the increase of thickness of the dip-coated CuO films its crystallite size increases and band gap decreases as evident from Fig. 7a and 7b, these could be attributed to the structural improvement of the CuO (see Table 1). The refractive index *(n)* can be used to calculate the bulk porosity of deposited CuO thin films using the following relation:[41]

$$Porosity = \left[1 - \left(\frac{n^2-1}{n_d^2-1}\right)\right] \times 100\% \quad (11)$$

Where $n_d$ is the refractive index of pore-free monoclinic CuO thin film ($n_d$ = 2.6)[42] and $n$ is the refractive index of deposited CuO thin film and the values of $n$ are taken at 500 nm wavelength region.

The refractive index of the films are evaluated using the following equation:[29]

$$n = \left(\frac{1+R}{1-R}\right) + \sqrt{\frac{4R}{(1-R)^2} - k^2} \quad (12)$$

Where $R$ is the reflectance, $k$ is the extinction coefficient, and $\lambda$ is the wavelength of the incident beam (see also Fig. S2 and Fig. S3 for details). The change in porosity of CuO thin films for various withdrawal speeds as a function of film thickness is displayed in Fig. 7c. It can be seen that the porosity of the films increases with the increase of film thickness, suggesting a lower packing density of thicker films. These results are also in good agreement with the surface morphology analyses (cf. Fig.4a - Fig. 4d). A film with high amount of bulk porosity is desirable for gas sensing purposes.[19]

### 3.3.2. Dielectric constants, Surface and Volume energy loss functions

The intrinsic property complex dielectric constant of a material modifies electromagnetic waves represent the excitation of the phonons and entire information of optical transition in a material. Dielectric constant can be calculated using the following relationship.[43]

$$\varepsilon = \varepsilon_1 + \varepsilon_2 = (n+ik)^2 \quad (13)$$

Where, $\varepsilon_1$ and $\varepsilon_2$ are real and complex part of dielectric constants and are given by,

$$\varepsilon_1 = n^2 - k^2 \quad (14)$$
$$\text{and } \varepsilon_2 = 2nk \quad (15)$$

The real part ($\varepsilon_1$) of the dielectric constant is associated with the dispersion or slowing down properties of the speed of light in the material while the imaginary part ($\varepsilon_2$) evaluates the absorption of energy from the electric field due to dipole motion. (Detail's graph of dielectric constants of dip-coated CuO can be found in Fig. S5 in the supplemental file). Figure 8a illustrates the variation of $\varepsilon_1$ and $\varepsilon_2$ of dip-coated CuO films as function of thickness. It is observed that both the value of $\varepsilon_1$ and $\varepsilon_2$ increases with the increase of film thickness may be due to improvement of the film crystallization and reduction of the atomic interplanar distance as well as minimization of the crystal defects. The same agreement of dielectric properties is found for CuO thin films prepared by CBD from other researchers.[44] The real and imaginary values are high for the film deposited with dip-speed 73 mm/s; it is due to high thickness as well as low band gap energy.

The volume energy loss function $(V_{ELF})$ and surface energy loss function $(S_{ELF})$ measure the probability of loss of energy of charge carriers while moving through the bulk and the surface of the material. The characteristic losses: $V_{ELF}$ and $S_{ELF}$ are related to dielectric properties of the material involved and can be determined by the following relations:[41]

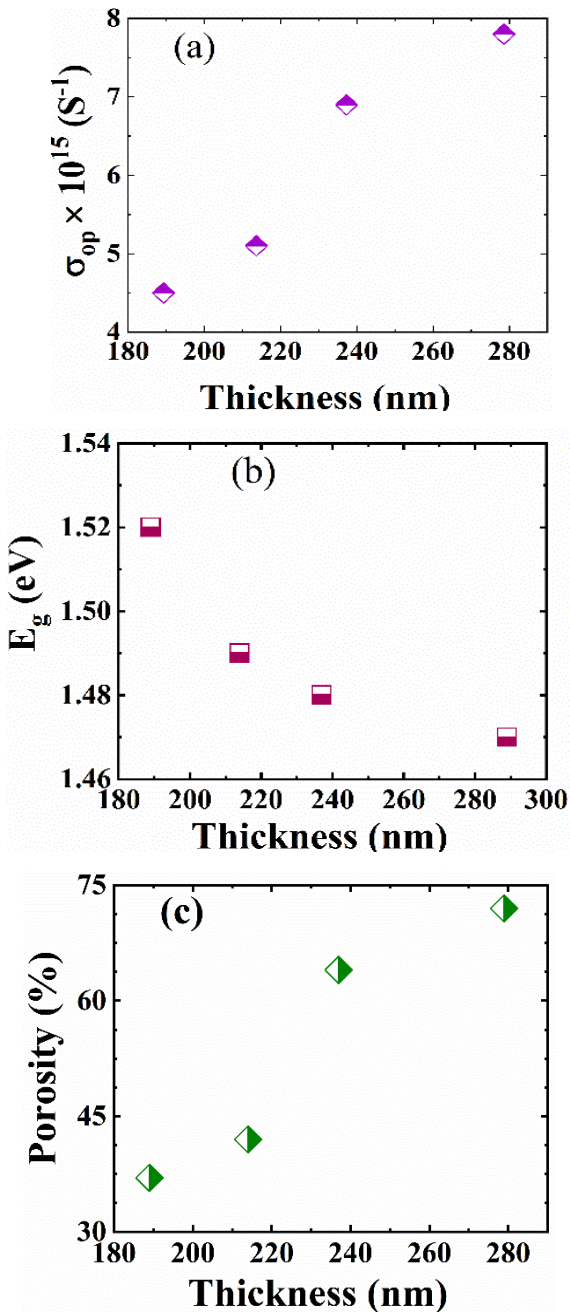

Fig. 7 (a) Optical conductivity (at $h\upsilon$ = 2.0 eV) (b) band gap energy and (c) change in porosity (at $\lambda$ = 500 nm) as a function of film thickness of CuO thin films grown on different withdrawal speeds.

$$V_{ELF} = \frac{\varepsilon_2}{\varepsilon_1^2 + \varepsilon_2^2} \quad (16)$$

$$S_{ELF} = \frac{\varepsilon_2}{[(\varepsilon_1^2+1)^2 + \varepsilon_2^2]} \quad (17)$$

The change in $V_{ELF}$ and $S_{ELF}$ (estimated at $h\upsilon$ = 2.0 eV; see also Fig. S6 for details) of dip-coated CuO films deposited at different withdrawal speeds shown in Fig. 8b as a function of film thickness. With the increase in film thickness the value of $V_{ELF}$ and $S_{ELF}$ decreases which indicating the decreased electron energy loss. Both $V_{ELF}$ and $S_{ELF}$ values were found to be the lowest for 0.73 mm/s dip-coated CuO film in this study.

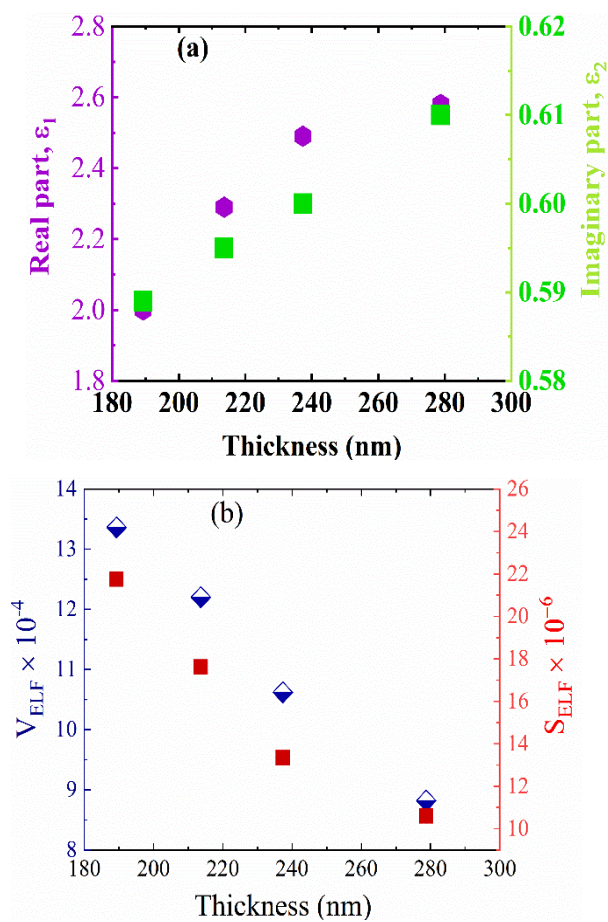

Fig. 8 (a) The real and imaginary part of dielectric constants and (b) Variation of $V_{ELF}$ and $S_{ELF}$ as a function of film thickness.

### 3.4. DC Electrical studies

The measurement of the lateral resistance through a thin square of material is sheet resistance. The sheet resistance of synthesized CuO thin films was measured by a four-point collinear probe method reported elsewhere[45] and shown in Fig. 9a. It can be seen that the

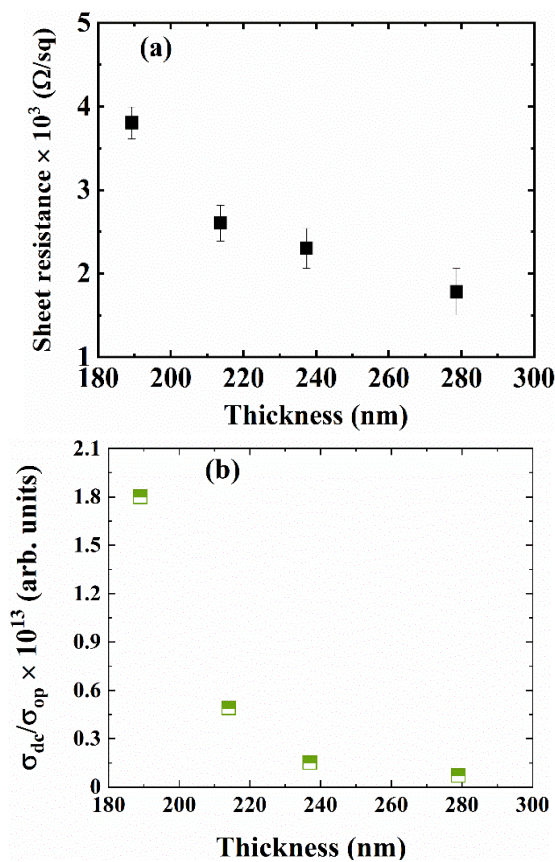

Fig. 9 (a) Variation of sheet resistance, $R_s$, and (b) the ratio of $\sigma_{dc}$ and $\sigma_{op}$ of CuO thin film as a function of film thickness.

sheet resistance of the dip-coated CuO films gradually decreases with decreasing withdrawal speeds (i.e., increasing film thickness) which can be attributed to the increment of crystallite size (see Table 1). In other words, with the increase of grain size, the grain boundary would decrease, which subsequently would decrease the scattering of charge carriers at the grain boundaries.[36] Hence, a decrease in the value of sheet resistance due to the increment of film thickness leads to the improvement of the film's electrical conductivity. Since the conductivity is positively correlated with the film thickness, improved crystallinity as well as decrement of optical bandgap, therefore, improvement of electrical conductivity may be attributed to the increase of carrier concentrations in the dip-coated films. The elemental composition of dip-coated CuO films investigated by WDXRF revealed that Cu-content tends to increase with the increase of withdrawal speeds. This would lead to increase of copper vacancy ($V_{cu}$) in the CuO matrix and results in enhance electrical conductivity of p-type CuO materials (see Table S1 and Fig. S8 in the supplemental file). Therefore, the higher carrier concentrations could extend the localized states near the band-edge(s) and reduces the band gap (see Table 2). The ratio of dc electrical conductivity ($\sigma_{dc}$) (sheet resistance converted to conductivity considering related film thickness) and optical conductivity ($\sigma_{op}$) versus film thickness is also shown in Fig. 9b. Previously we observed that $\sigma_{op}$ increases with increasing film thickness accompanied by the porosity increment in the dip-coated CuO films (cf. Fig. 7a and 7c). The value of $\sigma_{dc}/\sigma_{op}$ gradually decreases with the increase of film thickness. As mentioned above the ratio of $\sigma_{dc}/\sigma_{op}$ tend to fall off when the film thickness is higher. This type of transition can be attributed to the local uniformity of the deposited film as a function of thickness.[46] Therefore, a comparatively thicker film with porous morphology/structure would entail the high adsorption of gas species and a high electrical conductivity with lower energy bandgap may faster the gas detection response.

### 3.5. Wettability of CuO thin films

The humidity around the gas molecules and solid sensor surface can greatly affect the sensing performance of sensor. To this end, the wettability of dip-coated CuO thin film surfaces were estimated by measuring the water contact angle (WCA). A water droplet of ~10 μL was placed and allowed the droplet to stabilize for ~5 s on the CuO surface before WCA measurements. Figure 10 shows the water droplet profile and corresponding WCA on dip-coated CuO deposited by using 0.73, 1.75, 3.39, and 4.17 mm/s withdrawal speeds. It can be observed that the average WCA reduces from 98.78° to 65.25° with increasing withdrawal speeds. Among all dip-coated CuO films, the film prepared at 0.73 mm/s withdrawal speed exhibits the hydrophobicity nature accompanied by the highest WCA= 98.78°. The highest amount of porosity and the largest WCA of CuO surface dip-coated by using 0.73 mm/s withdrawal speed should exhibit better gas sensing response compared to the other films.[47] As a proof-of-concept, we employed some dip-coated CuO thin films/glass substrate to evaluate their $CO_2$ sensing performance at room temperature (RT) in the section 3.5.

### 3.6. $CO_2$ sensing properties

The gas-sensing performance of the dip-coated CuO thin film was investigated at RT using a home-made $CO_2$ sensing setup as shown in Fig. 11a. A high precision electrometer (Keithley 6517B) with spring-loaded contact pins in Kelvin probe technique was utilized to directly measure the I-V characteristic curves (with and without $CO_2$ gas) to show the variation of CuO films response with 10000 ppm of $CO_2$ gas and in absence of gas (Fig. 11b). It can be seen that the thicker films with higher porosity/porous surface morphology deposited with slower withdrawal speeds show a consistently higher response compared to thinner films. Besides, the detectable response is seen to be as low as 0.71 mA for the thinnest film. The difference of current values with applying voltage is mostly due to the $CO_2$ being adsorbed in the film and changing the bulk resistivity, rather than only on the surface nature.[7] Here, $CO_2$ gas has more chances to percolate through porous surface/bulk structure than the compact structure due to the availability of more active sites in the former, which may be reflected in Fig. 11b.

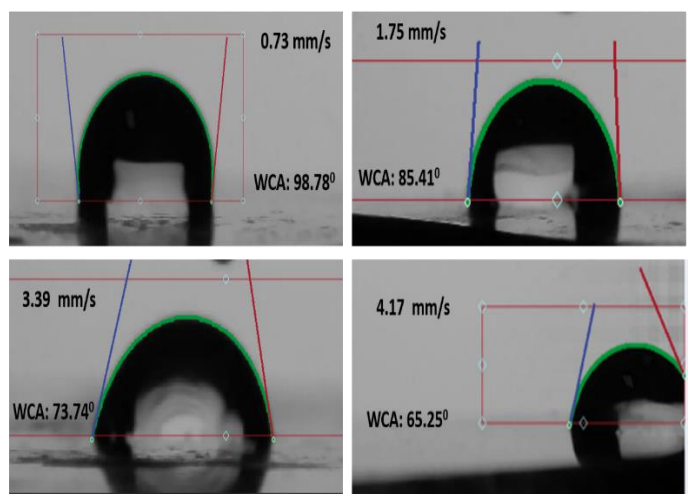

Fig. 10 The images of WCA measuring for dip-coated CuO thin films grown on SLG using 0.73, 1.75, 3.39, and 4.17 mm/s withdrawal speeds.

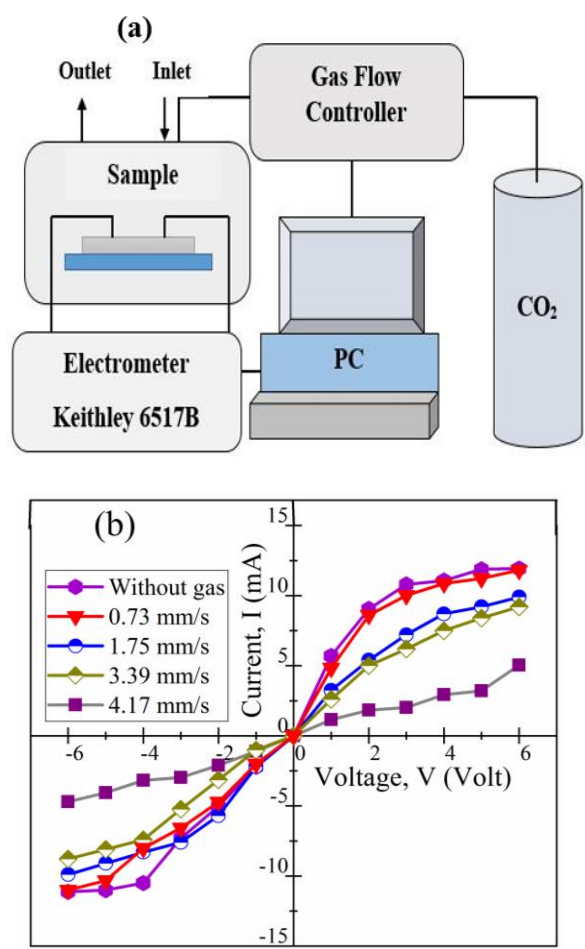

Fig. 11 (a) Schematic diagram of the homemade $CO_2$ gas sensitivity measurement setup used in this study. (b) I-V characteristic curve of without gas and in expose to 10000 ppm $CO_2$ for CuO thin films dip-coated with different withdrawal speeds.

The surface of the CuO thin film absorbs oxygen species ($O^{2-}$, $O^-$, and $O_2^-$) and diffuses through the pore. The adsorption of ionized species leads to the formation of hole-accumulation layers (HALs) on the surface of p-type CuO surface [48]. Upon contact with $CO_2$ gas, the resistance would decrease as the conduction occurs mainly along the near-surface HAL of p-type CuO therefore enhance the gas sensing response. [48]

The responsivity of the dip-coated CuO thin films with the variation of $CO_2$ gas concentrations was also studied and calculated by using the following relation:[29]

$$Response\ (\%) = \frac{R_a - R_g}{R_a} \times 100\% \qquad (18)$$

Here, $R_a$ is the resistance of the film in air and $R_g$ is the resistance upon exposure to $CO_2$ gas.

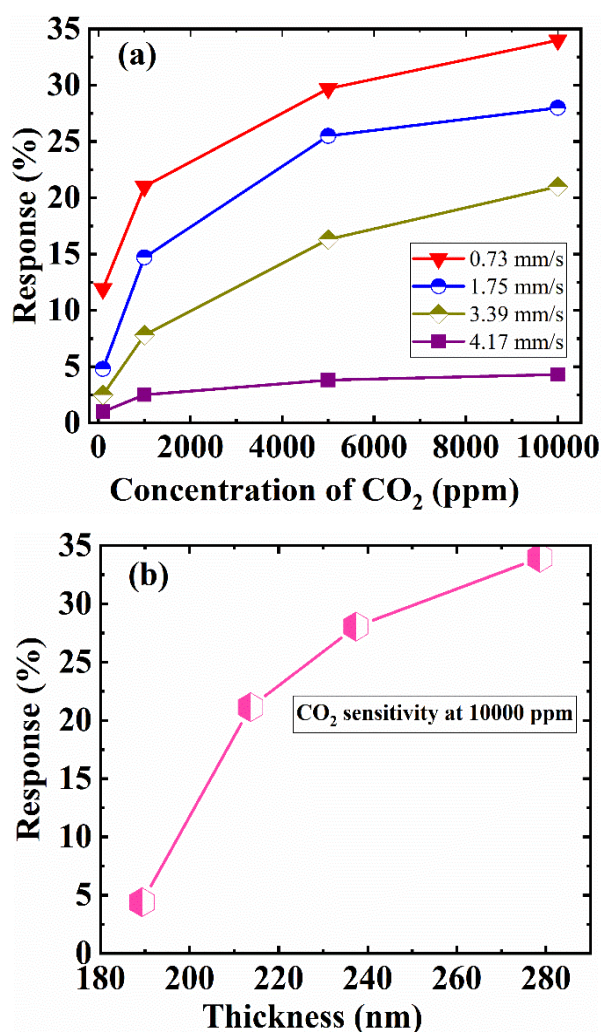

Fig. 12 (a) Response characteristics of the dip-coated CuO thin films as a function of various $CO_2$ gas concentration with different withdrawal speeds namely 0.73, 1.75, 3.39, and 4.17 mm/s, and (b) response characteristics as a function of film thickness in presence of 10000 ppm $CO_2$ gas.

Figure 12a shows the response characteristics of CuO thin films as a function of $CO_2$ concentrations The consistent increase of the $CO_2$ response with increasing film thickness shown in Fig. 12b can be attributed to the combined effects of the improved crystallinity, better opto-electrical conductivity, increased porous and hydrophobic nature of the said film. Therefore, the maximum response to $CO_2$ vapor for the sample prepared at 0.73 mm/s withdrawal speeds could be attributed to its highest porous structure, hydrophobicity nature as well as opto-electrical conductivity among the all dip-coated CuO films studied. Similar results have also been reported in the literature.[44] Besides, the observation from SEM micrographs of these samples shows that thicker films have much larger grains than thinner ones. Larger grain sizes provide lower resistive pathways for charge carriers accompanied by higher gas sensing response. These preliminary gas sensing observations at room temperature suggest that dip-coated CuO thin films could be utilized for $CO_2$ detection. Further investigations on preparing CuO-based composite by dip-coating are currently in progress to improve the sensing/optoelectronic properties and will be communicated elsewhere.

## 4. Conclusion

The impact of film thickness manifested by dip-coating withdrawal speeds on the structural, morphological, opto-electrical, and wettability nature of the dip-coated CuO thin films were investigated in detail. The structural properties characterized by XRD revealed that the thin films are polycrystalline in nature and have monoclinic CuO crystal structure with preferential orientations along the (111) and ($\bar{1}11$) planes and they are free from impurity phases further confirmed by Raman spectroscopic analysis. The crystallite size of the deposited films was found to increase from 15.41 to 21.83 nm with increasing film thickness. The porous morphology of CuO surface were seen to be increasing with the increase of film thickness. The band gap of dip-coated CuO films were reduced from 1.52 to 1.47 eV accompanied by the expansion of the Urbach band tails near band edge due to the increase of carrier concentration. With the increment of film thickness, optical, electrical conductivity as well as the dielectric constants were found to be increased notably due to their increment of crystallite as well as particle size. The wettability of the dip-coated CuO surface was found to be decreasing with the increase of film thickness as well as porosity. The sample deposited to 0.73 mm/s withdrawal speed showed the maximum response (~34%) at 10000 ppm of $CO_2$ vapor in the air. This study suggests that the dip-coated CuO thin films with tailorable physical properties can offer a great potential for $CO_2$ gas sensing applications.

## Author Contributions

**A.M.M. Musa:** Investigation, Formal analysis, Writing - original draft. **S.F.U. Farhad:** Conceptualization, Investigation, Methodology, Formal analysis, Project administration, Supervision, Funding acquisition, Writing - original draft. **M.A. Gafur:** Methodology,

Visualization, Resources, Supervision. **A.T.M.K. Jamil:** Conceptualization, Validation, Formal analysis, Supervision, Writing - review & editing.

## Conflicts of interest

The authors declare no competing financial interest.

## Acknowledgements

The authors would like to concede the help of Bangladesh Council of Scientific and Industrial Research (BCSIR), Dhaka-1205, Bangladesh for providing the laboratory facilities. S.F.U. Farhad and M.A. Gafur thank Ms. Julia Khanum, SSO, BCSIR for helping with WDXRF experiment. A.M.M. Musa and A.T.M.K. Jamil gratefully acknowledge the Dhaka University of Engineering and Technology, Gazipur-1707, Bangladesh for the financial support for this research. S.F.U. Farhad acknowledges support of TWAS grant# 20-143 RG/PHYS/AS I for Industrial Physics Division, BCSIR.

## Notes and references

# Effects of withdrawal speeds on the structural, morphological, electrical, and optical properties of CuO thin films synthesized by dip-coating for $CO_2$ gas sensing

A.M.M. Musa[a,b,‡], S.F.U. Farhaa[a,c,*,‡], M.A. Gafur[a], and A.T.M.K. Jamil[b]

[a] Bangladesh Council of Scientific and Industrial Research (BCSIR), Dhaka 1205, Bangladesh

[b] Department of Physics, Dhaka University of Engineering & Technology (DUET), Gazipur-1707, Bangladesh

[c] Central Analytical and Research Facilities (CARF), BCSIR, Dhaka 1205, Bangladesh

[‡]Equal contribution

*Correspondence: sf1878@my.bristol.ac.uk ; s.f.u.farhad@bcsir.gov.bd

## The transmission spectra and absorption coefficient:

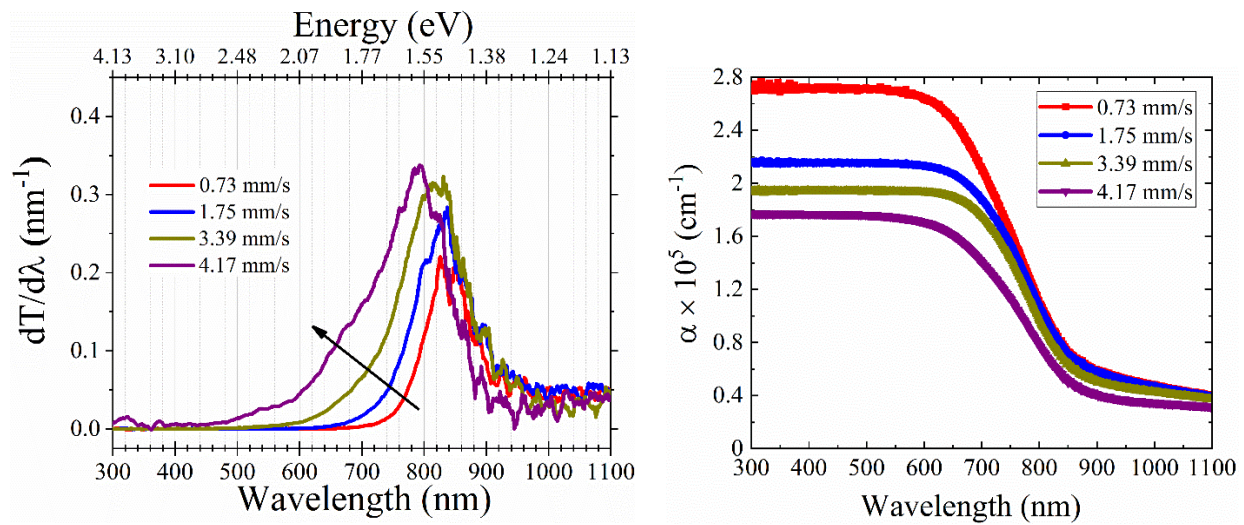

**Fig. S1:** First derivative of transmission spectra (left) and absorption coefficient *(α)* (right) as a function of wavelength of 0.35 M CuO thin films deposited at various withdrawal speeds.

The arrow mark in the right figure indicating the increase of band tails with the increase of withdrawal speeds. The inflection points of *dT/dλ* showing that the optical band gap is in the range of ~1.44-1.57 eV (i.e., 790 – 860 nm wavelength which also corroborates the diffuse reflection spectra of these samples as shown in Fig. S2 below). The graph of *α* versus wavelength is shown in Fig S1 (right). The value of *α* depends on film thickness.[1] The higher value of *α* of the deposited film is found for lower withdrawal speed 73 mm/s. It may be due to the increment of photon absorption which leads to multiple light scattering at grain boundaries.

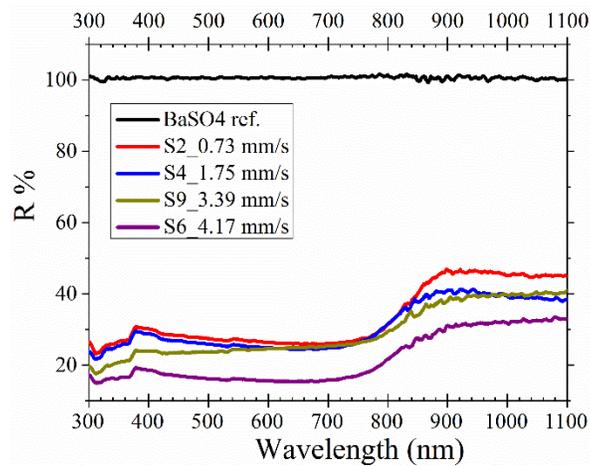

**Fig. S2**: Diffuse reflection spectra of deposited dip-coated CuO thin films for different withdrawal speeds.

## The refractive index and porosity (%):

The refractive index as a function of wavelength is estimated from diffuse reflection data of the deposited films is shown in Fig. S3a. Moreover, porosity (%) versus wavelength of CuO thin films is shown in Fig. S3b. The rising value of porosity is inverse with the refractive index of the film.[2] The lower *n* value is observed when film thickness is high which exhibited the higher porosity of the deposited sample.

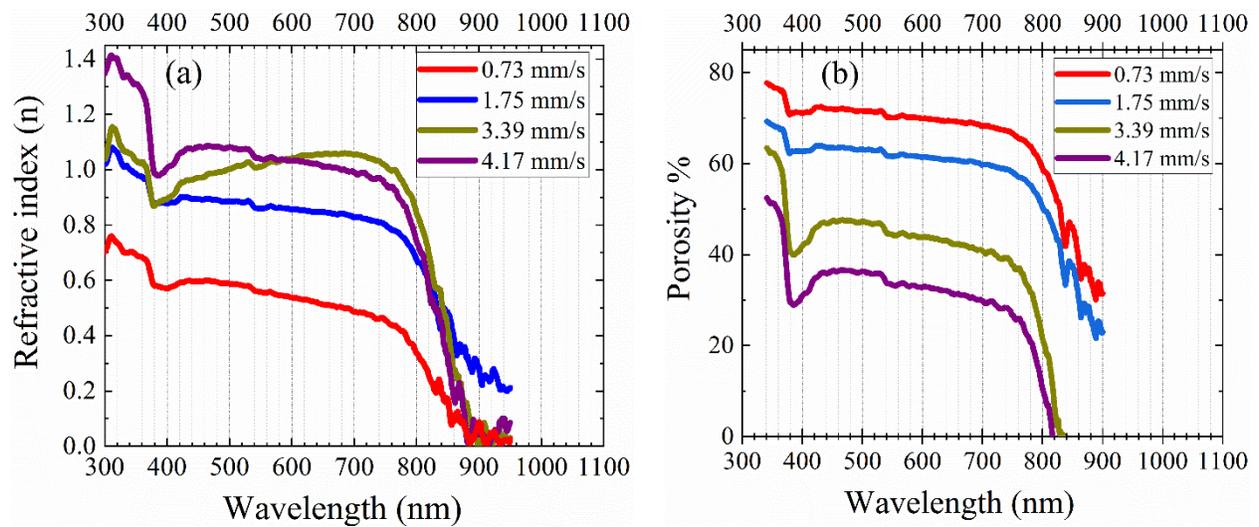

**Fig. S3: (a)** The refractive index *(n)* corresponding to diffuse reflection data (Fig. S2) and **(b)** porosity (%) as a function of wavelength of 0.35 M CuO thin films deposited at various withdrawal speed.

## Urbach energy:

Urbach energy $(E_u)$ of CuO thin film is calculated from the absorption coefficient $(\alpha)$ data of deposited films. Fig. S4 represent the value of $E_u$ from the area marked by the rectangle. It is observed that, with the increase of film thickness the value of $E_g$ decreases while the value of $E_u$ increases.[3] However, the lowest $E_g$ and highest $E_u$ is found for the film deposited at 0.73 mm/s withdrawal speed.

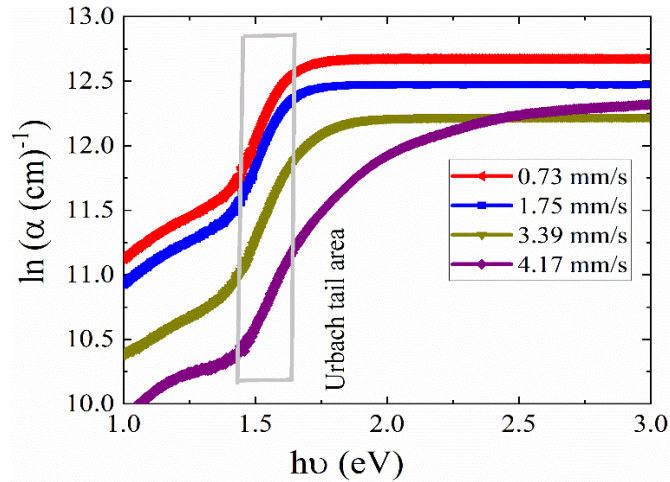

**Fig. S4:** Urbach tail $(E_u)$ estimation (area marked by the rectangle) from the $ln\alpha$ vs $hv$ curves of CuO thin films by dip-coating at various withdrawal speeds.

## Real and imaginary part of dielectric constants:

The real part $(\varepsilon_1)$ and imaginary part $(\varepsilon_2)$ of dielectric constants as a function of wavelength are shown in Fig S5a and S5b. It is seen that both $\varepsilon_1$ and $\varepsilon_2$ are increases for slower withdrawal speed. The dielectric constants values are slowly decreasing with the increase of wavelength of the deposited CuO thin films.

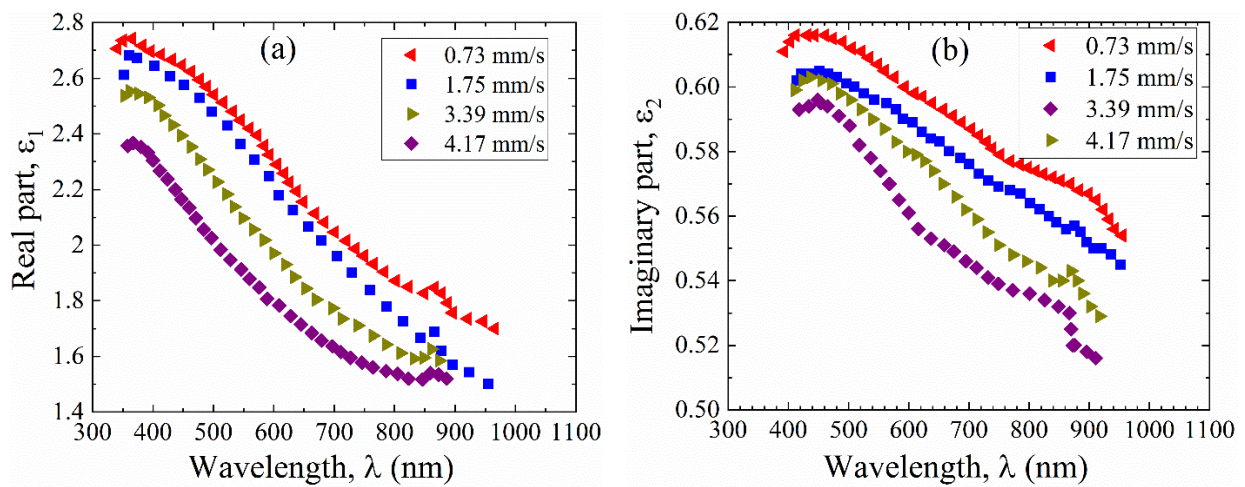

**Fig. S5:** The (a) real and (b) imaginary part of dielectric constants as a function of wavelength of CuO thin films prepared by different withdrawal speeds.

## The volume and surface energy loss function and optical conductivity:

The volume energy loss function *(V$_{ELF}$)* and surface energy loss function *(S$_{ELF}$)* as a function of *hυ* are shown in Fig. S6a and S6b. With the increase in film thickness the value of *V$_{ELF}$* and *S$_{ELF}$* decreases which indicating the decreased electron energy loss.[2] The optical conductivity *(σ$_{op}$)* versus *hυ* graph shown in Fig. S7. The value of *σ$_{op}$* rises with the increase in film thickness.[4]

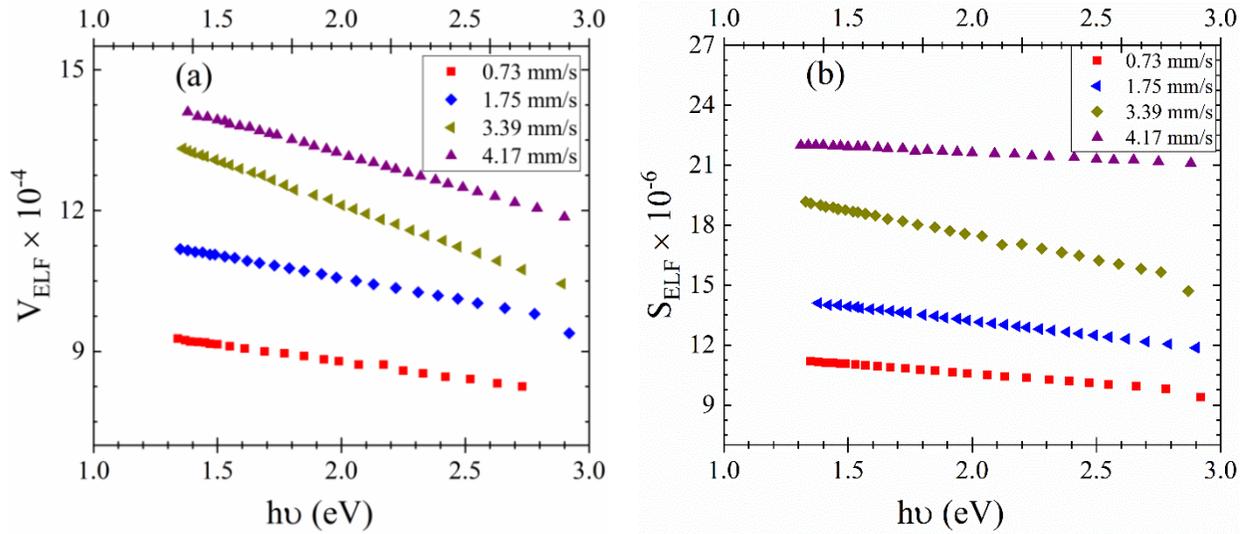

**Fig. S6:** The (a) volume energy loss function (V$_{ELF}$) and (b) surface energy loss function (S$_{ELF}$) as a function of wavelength of CuO thin films prepared by different withdrawal speeds.

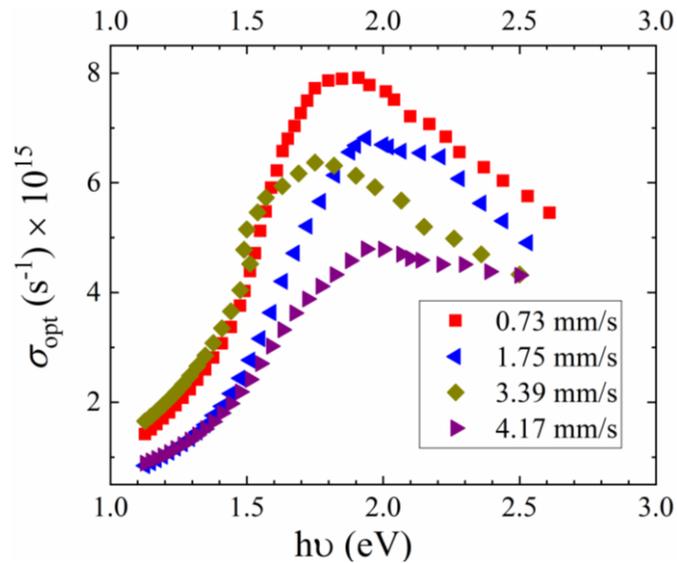

**Fig. S7:** Optical conductivity as a function of the photon energy of CuO thin film for different withdrawal speeds.

**Table S1.** Semi-quantitative elemental composition analyses by wavelength dispersive X-ray fluorescence (WDXRF) spectrometer of thin films (analyzing sample area ~10 mm) and CuO pellet (purity~99.95%; sampling area ~30 mm). The thickness, electrical sheet resistance, type of conductivity of dip-coated CuO films with the variation of withdrawal speeds are also included for comparison purposes. Hot point probe (thermoelectric) measurements were conducted at a temperature difference of ~150 $^0$C between the two probe of ~1 cm apart on the CuO film surface and measurement values were taken after allowing the voltage readings to stabilize for ~30 s. Heat source (soldering iron) was placed on the positive lead of the Digital Multimeter (Model: ANENG Q1). All dip-coated CuO films exhibited a p-type conductivity (see far right column of the table below).

| Withdrawal speed (mm/s) | Thickness (nm) ± 5 | Electrical Sheet resistance (×10$^3$ Ω/Sqr.) | Cu-Kα Peak intensity (a.u.) | CuO (%) (estimated using WDXRF Software Calc.) | Hot Probe Voltage reading (mV) ± 5 |
|---|---|---|---|---|---|
| 0.73 | 279 | 1.85 | 12.40 | 4.10 | -35 |
| 1.75 | 237 | 2.15 | 15.67 | 4.98 | -41 |
| 3.39 | 214 | 2.50 | 21.60 | 8.73 | -51 |
| 4.17 | 189 | 3.85 | 36.17 | 27.60 | -50 |
| CuO pellet (purity ~99.95%) | 2 mm | - | 8288 | 98.10 | - |

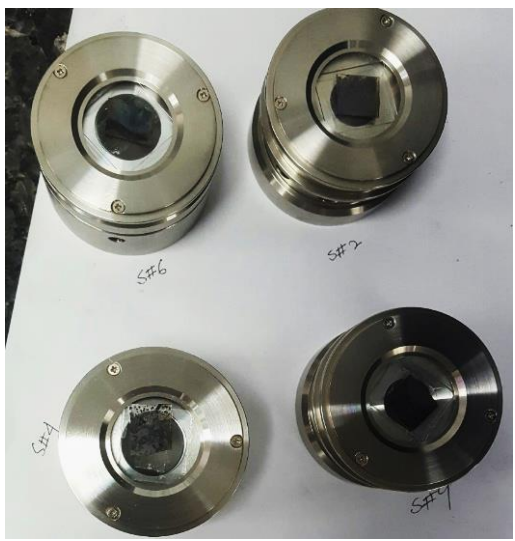

**Fig. S8:** Photograph of dip-coated CuO thin films on WDXRF sample holders used for semi-quantitative elemental composition analyses. The 10 mm diaphragm was used for confined sample area to avoid signal from the sample holder etc. (S2:0.73 mm/s; S4:1.75 mm/s; S9:3.39 mm/s; S6: 4.17 mm/s)